\documentclass[11pt]{article}
\usepackage{moriond,epsfig}
\usepackage{graphicx}

\def\be{\begin{equation}}
\def\ee{\end{equation}}
\def\bea{\begin{eqnarray}}
\def\eea{\end{eqnarray}}

\begin{document}
\vspace*{4cm}
\title{CHARM PRODUCTION AT RHIC}

\author{ Ming X. Liu }
\address{MS H846, P-25, Los Alamos National Laboratory \\ 
  Los Alamos, NM 87545, U.S.A.}

\maketitle

\abstracts{ 
The latest results for open charm and $J/\psi$ production 
in p-p, d-Au and Au-Au from the PHENIX and STAR experiments  
at $\sqrt{s_{NN}}=200$ GeV at RHIC are presented. 
The preliminary data show open charm production follows
binary scaling in d-Au and Au-Au collisions at RHIC. In d-Au collisions, 
a suppression in $J/\psi$ production 
has been observed at the forward rapidity (d direction), 
at the backward rapidity (Au direction), $J/\psi$ production
seems strongly dependent on collisions centrality. 
The implications of heavy flavor production in cold (d-Au) 
and hot (Au-Au) nuclear media at RHIC are discussed.
}

\section{Introduction}
Heavy flavor particles provide an important tool for studying the nuclear medium created in 
heavy ion collisions. 
The suppression of heavy quarkonium production in high energy heavy ion collisions 
is predicted as one of the signals for a phase transition of nuclear matter from 
confined to deconfined quarks and gluons, the so called quark-gluon plasma (QGP). 
However, other competing nuclear effects such as parton shadowing, 
heavy quark energy loss, and charm recombination will also affect the 
overall charmonium production. The hot nuclear medium created
in heavy ion collisions
could also increase charm production by opening up more phase space,~\cite{OPEN-charm} and therefore 
could affect both open charm and charmonium production.  

A strong suppression of high $p_{T}$ light hadron production
has been observed at RHIC in Au-Au collisions,~\cite{HADRON-sup} which is most likely due to parton 
energy loss in the hot and dense nuclear medium (or QGP); for heavy quarks, 
this effect may be reduced due to the dead cone effect,~\cite{HEAVY-sup} but this has not been 
confirmed by experimental data. 
The recent results for particle production at large forward and backward rapidity 
in d-Au collisions from the PHENIX experiment show that nuclear medium
effects could play a significant role in the interpretation of the AuAu data at a large 
rapidity~\cite{MUON-RCP}. 
Therefore, it is very important to systematically measure open charm and $J/\psi$ production in p-p, 
d-Au and Au-Au collisions to fully understand the underlying physics.

\begin{figure}
  \vskip 1.5cm
  \begin{minipage}{7.5cm}
    \hspace{1.0cm}\includegraphics[width=6.0cm,height=5.0cm]{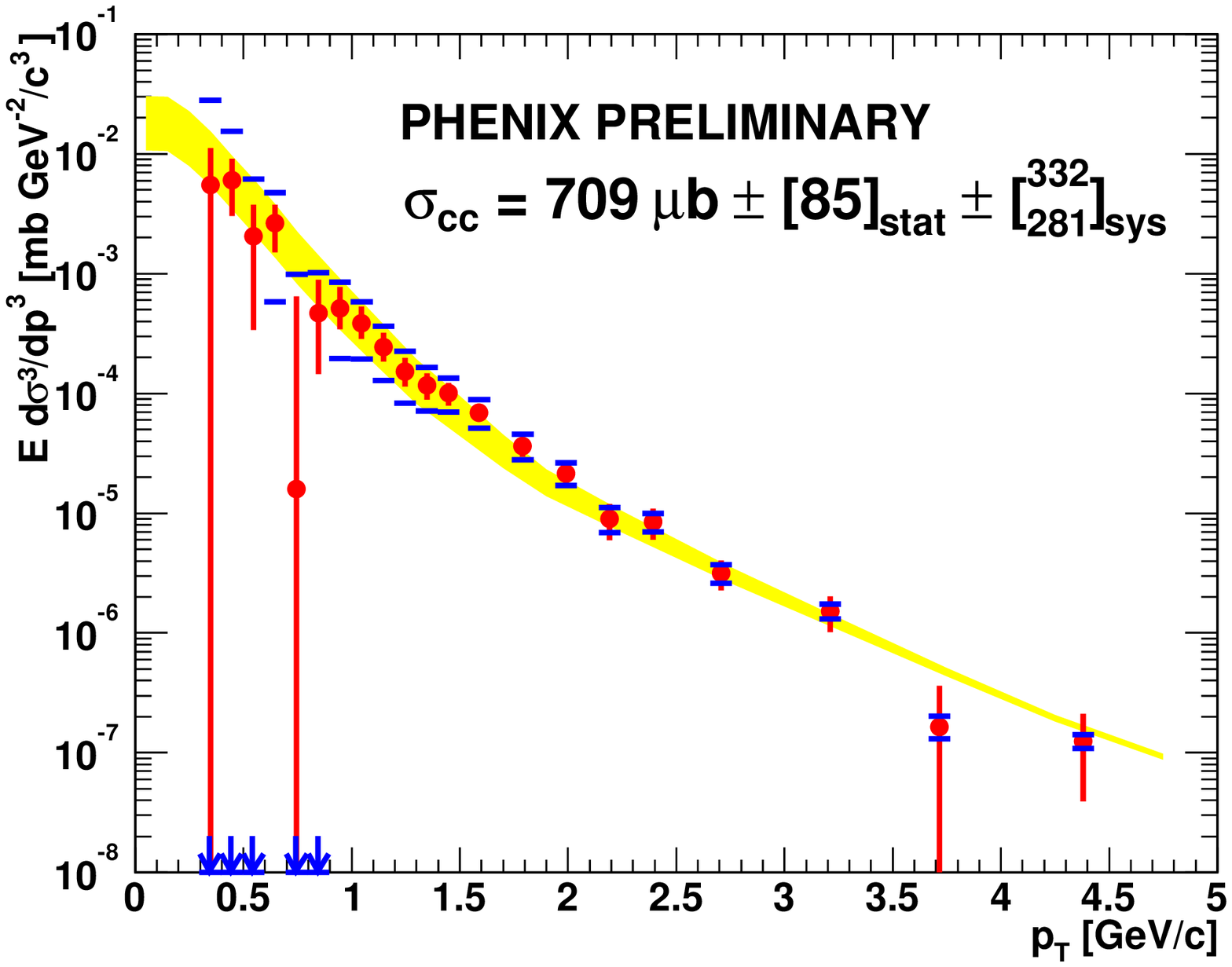}
    \caption{Non-photonic electron $p_{T}$ spectrum from p-p collisions.}
    \label{fig:PHENIX-Charm}
  \end{minipage}
  \begin{minipage}{7.5cm} 
    \hspace{1.0cm}\includegraphics[width=6.0cm,height=5.0cm]{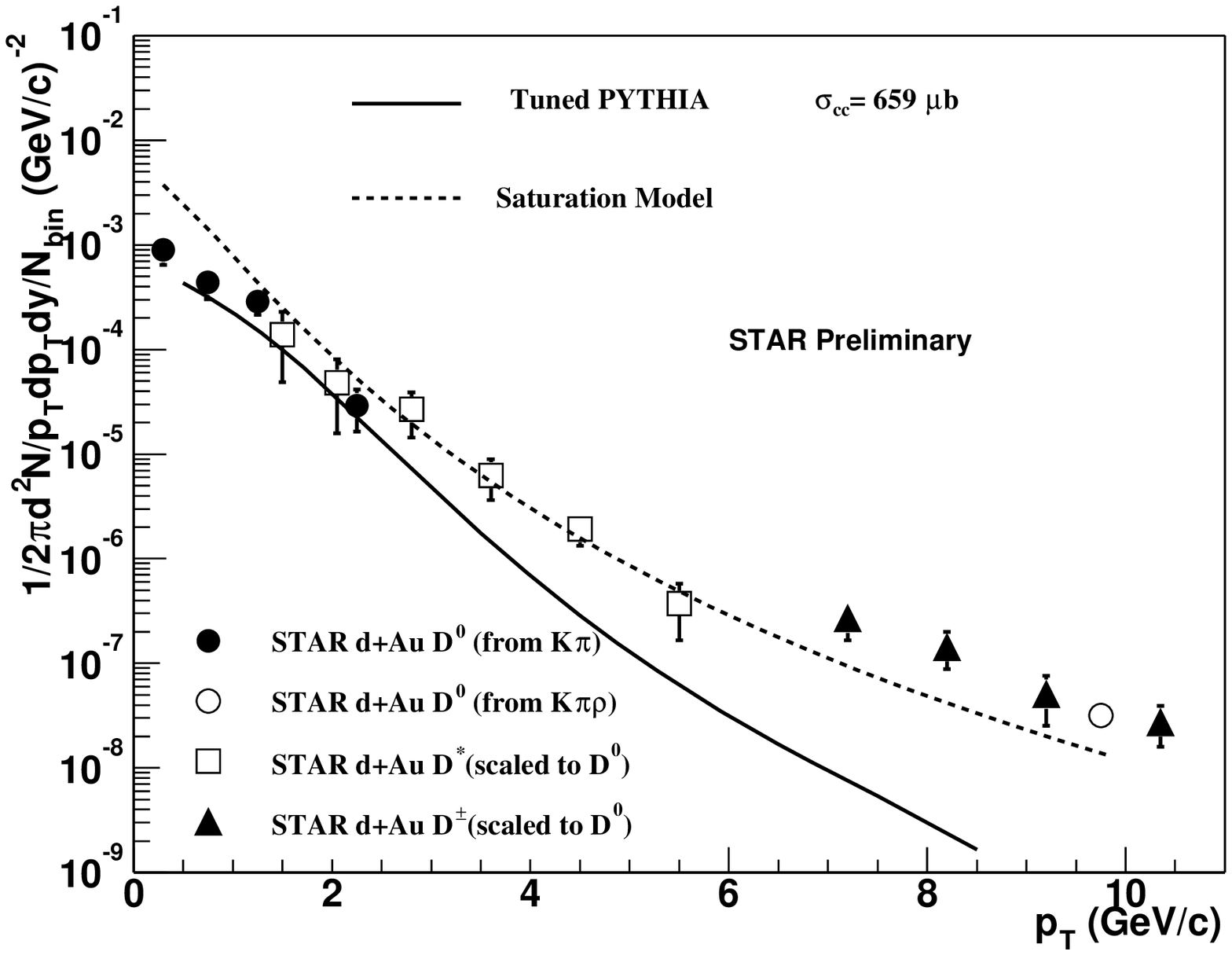}
    \caption{\label{fig:STAR-Charm}$D$ meson $p_{T}$ spectrum in d-Au collisions.}
  \end{minipage}
\end{figure}

\section{Open charm production}
At RHIC energies, open charm and $J/\psi$ particles are predominantly produced 
through gluon-gluon fusion processes.
Study of open charm production in p-p and d-Au collisions helps us to understand
the normal nuclear effects, such as gluon (anti)shadowing and particle energy 
loss in cold nuclear medium. Open charm production in p-p and d-Au collisions 
also serves as a baseline for Au-Au system where hot and dense nuclear medium is expected to be created.

The PHENIX experiment has measured single electron spectra in p-p, d-Au and Au-Au 
collisions at central rapidity $|\eta|<0.35$.~\cite{PHENIX-ele} After subtracting the photonic 
contributions (photon conversion, Dalitz and other light hadron decays), 
the remaining electrons are mostly from open charm
and open beauty semileptonic decays. Figure~\ref{fig:PHENIX-Charm} shows the resulting 
single electron $p_{T}$ spectrum in p-p  collisions at $\sqrt{s}=200$ GeV, from which 
the open charm total cross section is  extracted, 
$\sigma^{pp}_{c\bar{c}} = 709 \pm 85(stat) \pm ^{332}_{281} (sys) \mu b$.    

The STAR experiment, on the other hand, has measured not only the inclusive electron spectra but also
fully reconstructed the $D^{0,\pm}$ and $D^{*}$ mesons at central rapidity $|\eta|<0.5$ 
in hadronic channels, such as $D^{0} \rightarrow \pi^{-} K^{+}$. 
This is the first direct measurement of open charm production at RHIC.~\cite{STAR-D} 
Figure~\ref{fig:STAR-Charm} shows the reconstructed $D$ meson $p_{T}$ spectrum in d-Au 
collisions at $\sqrt{s_{NN}}=200$ GeV. The scaled nucleon-nucleon open charm cross 
section is given by 
$\sigma^{NN}_{c\bar{c}} = 1.12 \pm 0.20(stat) \pm 0.37(sys)  mb$, which is in a good 
agreement with the PHENIX results.

\begin{figure}
  \begin{center}
    \vskip 2.5cm
    \hskip 1.0cm \epsfig{figure=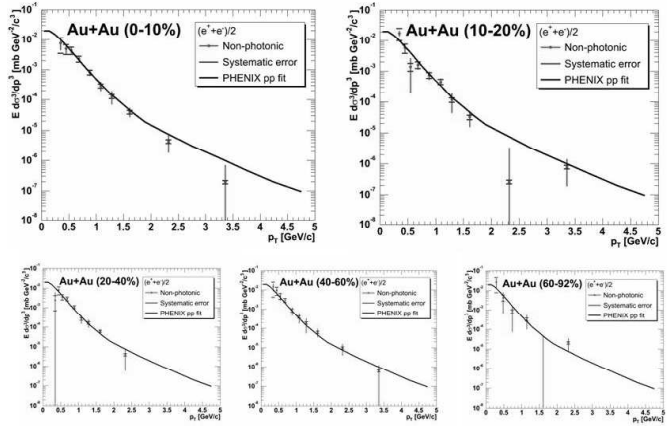,height=8.0cm}
    \vskip -2.5cm
    \caption{Electron $p_{T}$ spectra from 5 centrality classes in Au-Au collisions.}
    \label{fig:CharmAA-ele}
  \end{center}
\end{figure}

\begin{figure}
  \begin{minipage}{7.5cm}
    \vskip 0.0cm
    \hspace{1.0cm}\includegraphics[width=8.0cm,height=6.0cm]{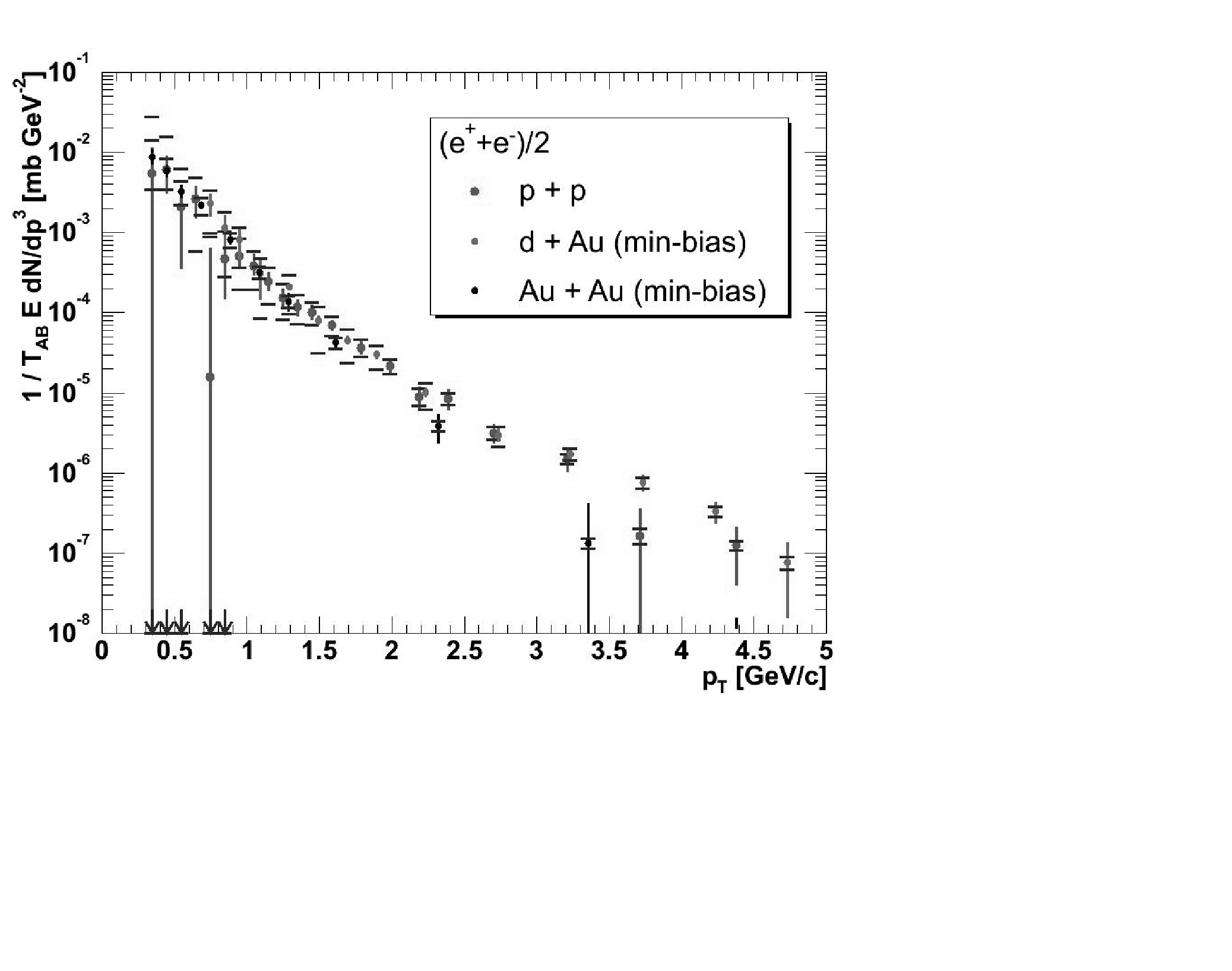}
    \vskip -2.0cm
    \caption{$<N_{coll}>$ scaled electron $p_{T}$ spectra from p-p, d-Au and Au-Au collisions,
      from PHENIX.}
    \label{fig:CharmAA-scale}
  \end{minipage}
  \begin{minipage}{7.5cm} 
    \hspace{1.0cm}\includegraphics[width=6.0cm,height=4.5cm]{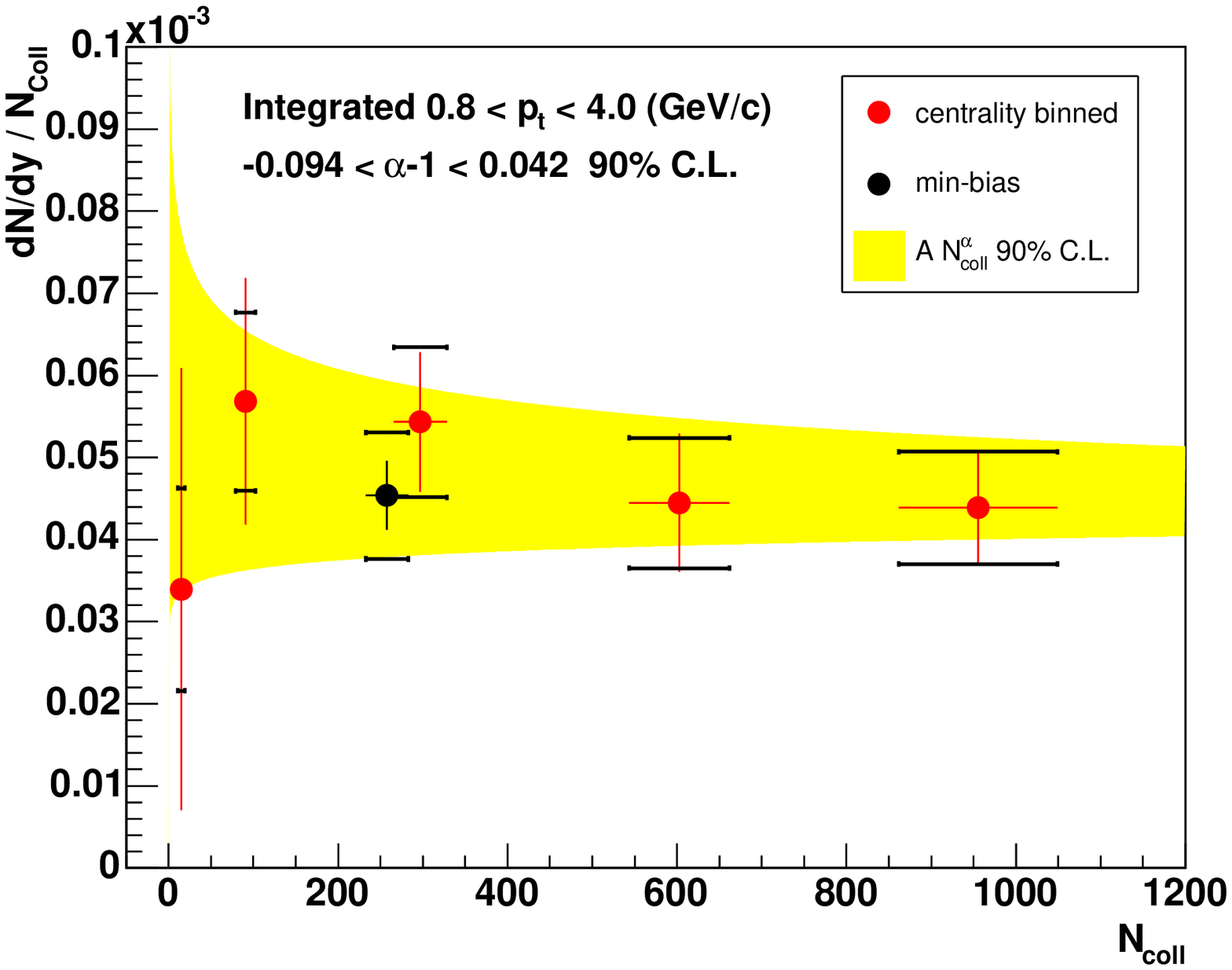}
    \caption{\label{fig:CharmAA} Integrated electron $dN/dy /<N_{coll}>$ vs $<N_{coll}>$
      in Au-Au collisions, from PHENIX.}
  \end{minipage}
\end{figure}

The effect of the nuclear medium on open charm production 
is studied by looking at (non-photonic) electron production as a function of collision 
centrality which is related to the impact parameter of two colliding nuclei. 
Figure~\ref{fig:CharmAA-ele} show the electron $p_{T}$ spectra in 5 different 
centrality bins in Au-Au collisions, fitted to the electron $p_{T}$ shape from p-p data.    
Figure~\ref{fig:CharmAA-scale} show the $<N_{coll}>$ scaled electron $p_{T}$ 
spectra from minimum biased p-p, d-Au and Au-Au data from the PHENIX, 
where $<N_{coll}>$, the average number of binary 
collisions, is estimated through a Glauber model calculation.
Figure~\ref{fig:CharmAA} shows the $<N_{coll}>$ scaled single electron yield in Au-Au collision 
as a function of $<N_{coll}>$, the data points are integrated 
over $0.8<p_{T}<4.0$ GeV and $|\eta| < 0.35$ at each centrality bin.
Within the experimental error, open charm production in d-Au and Au-Au system agrees with 
the binary scaling hypothesis. Thus it appears that there is no strong nuclear dependence 
of open charm production at RHIC.

\section{$J/\psi$ production}
$J/\psi$ production is predicted to be sensitive to the creation 
of the QGP in heavy ion collisions and an anomalous suppression has been observed 
in Pb-Pb collisions at CERN~\cite{NA50}. It is very important to check 
$J/\psi$ production in both the cold and hot nuclear media at RHIC.

The PHENIX experiment has measured $J/\psi$ 
in both p-p and d-Au collisions through the dielectron and dimuon channels, with good statistics in the 
rapidity range $|\eta|<0.35$ and $1.1 < |\eta| < 2.2$,~\cite{PHENIX-JPsi} and 
expects to detect a few thousand of $J/\psi$ particles from the
high luminosity Au-Au run that just finished at RHIC this year.

\begin{figure}
  \vskip 2.2cm
  \begin{minipage}{7.5cm}
    \hspace{1.0cm}\includegraphics[width=5.0cm,height=3.8cm]{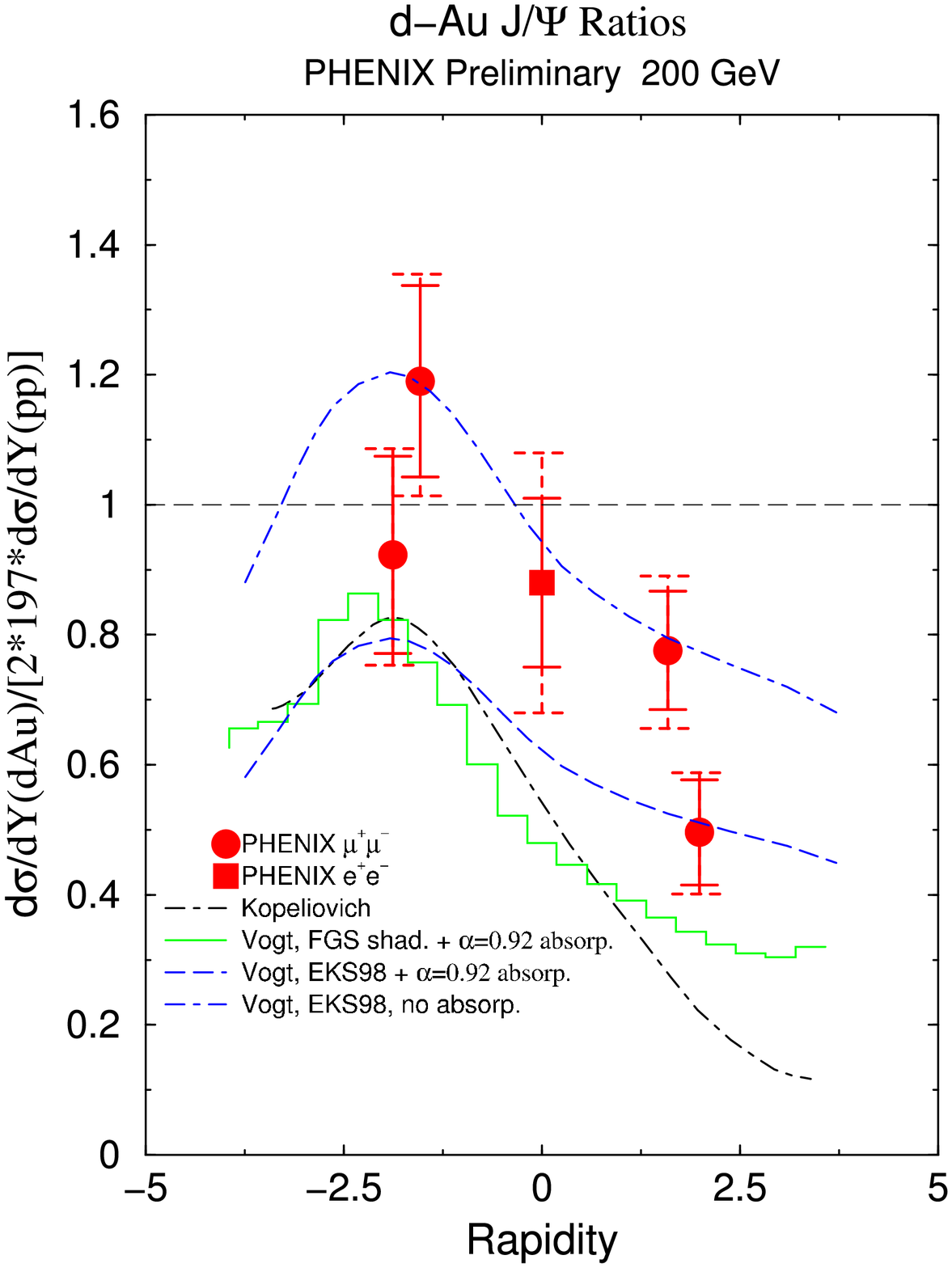}
    \caption{Ratio of d-Au  to p-p $J/\psi$ differential cross section versus rapidity, 
      normalized by $2\times197$. }
    \label{fig:JPSI-y}
  \end{minipage}
  \begin{minipage}{7.5cm}
    \hspace{1.0cm}\vspace{3.0cm}\includegraphics[width=3.8cm,height=3.7cm]{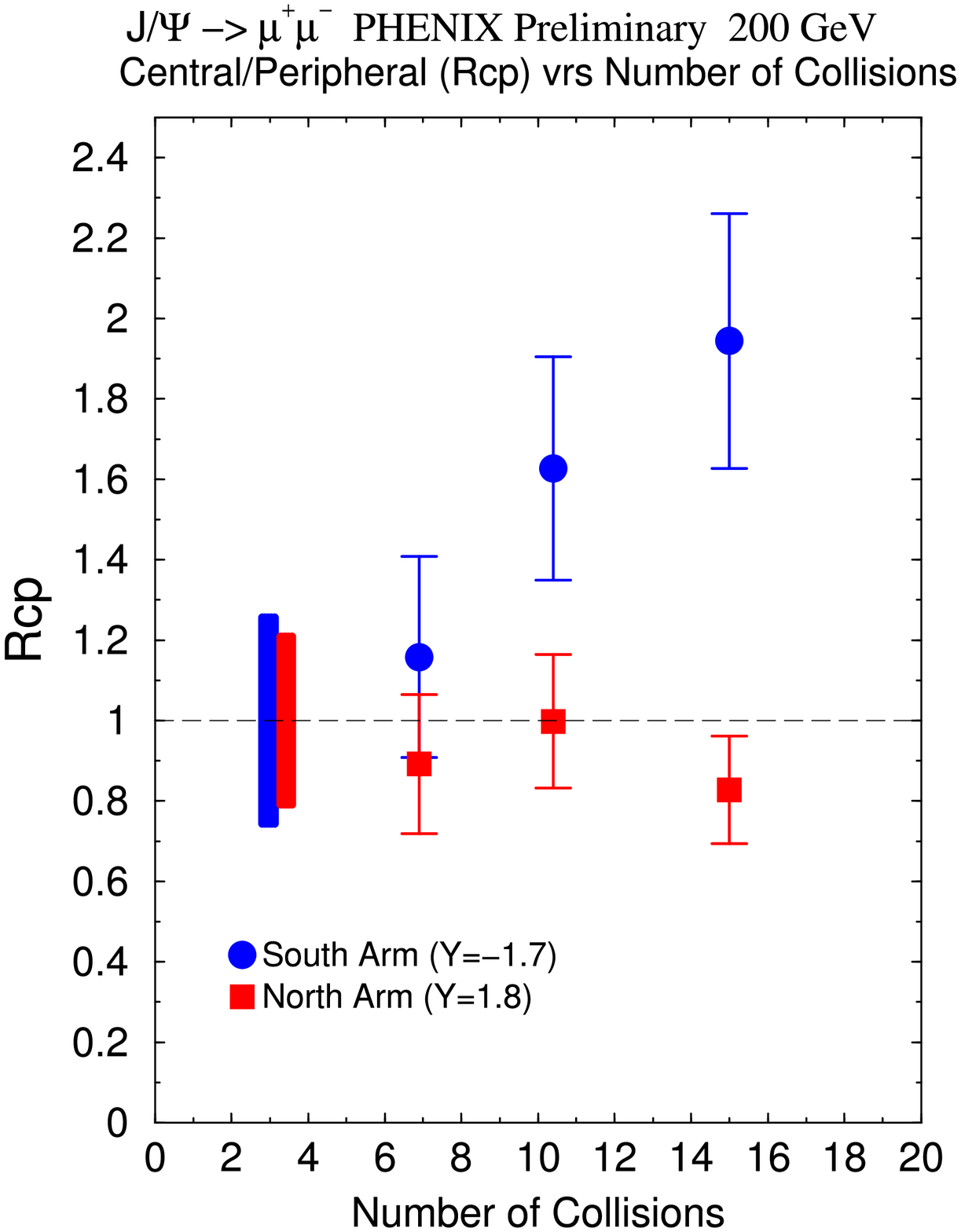}
    \vskip -2.5cm
    \caption{\label{fig:RCP} Centrality dependent $J/\psi$  $R_{CP}$ versus the average 
      number of binary collisions in d-Au collisions. }
  \end{minipage}
\end{figure}

The effects of cold nuclear medium on $J/\psi$ production are studied in d-Au collisions
by measuring the nuclear modification factor 
$R_{dA} = \frac{1}{2\times197}\frac{d\sigma^{dAu}/dy}{d\sigma^{pp}/dy}$,
which is the ratio of the $J/\psi$ yields observed in d-Au collisions relative to p-p system, 
scaled by $2 \times 197$.  Figure~\ref{fig:JPSI-y} shows $R_{dA}$ as a function of rapidity $y$.
While this ratio is close to unity 
at $y = 0$, a suppression has been 
observed at the forward rapidity, indicating 
possible gluon shadowing in Au nuclei.
Also plotted there are several theoretical model 
calculations~\cite{Vogot-Boris} with different assumptions about 
gluon shadowing and energy loss. 
Due to limited statistics of the current data, it is difficult to quantify 
the contributions from various processes.

The centrality dependence of $J/\psi$ production in d-Au collisions  has also been studied by PHENIX 
at both forward and backward rapidities. 
$R_{CP}$, the ratio of $J/\psi$ yields normalized by $<N_{coll}>$ in 
each centrality bin relative to the most peripheral one, 
is shown in  Figure~\ref{fig:RCP}. The preliminary data  show a strong dependence 
of $J/\psi$ production on collision centrality in the backward rapidity.

\section*{Conclusions}
Open charm and $J/\psi$ production have been measured by the PHENIX and 
STAR experiments in p-p, d-Au and Au-Au
collisions at RHIC. A high statistic $J/\psi$ measurement is expected from the recent 
high luminosity Au-Au run at RHIC.  
Within experimental errors, open charm production in d-Au and Au-Au collisions
is consistent with a binary  scaling hypothesis, indicating that there is no strong 
charm enhancement or suppression in heavy ion collisions at RHIC. 
From $J/\psi$ measurements in p-p and d-Au systems, possible shadowing of gluon distribution 
functions have been observed at large forward rapidity in d-Au collisions, however, due to 
limited statistics of current data, we can't distinguish various theoretical models. 
An interesting observation is the strong  centrality dependence of $J/\psi$ production 
at the backward rapidity in d-Au collisions, as measured by PHENIX. Theoretical work to understand 
what have been observed in charm production at RHIC is in progress.

\end{document}